\newcommand{\Msun}{~M_\odot}
\newcommand{\msun}{M_\odot}

\newcommand{\kms}{\rm ~km~s^{-1}}
\newcommand{\ergs}{\rm ~erg~s^{-1}}

\newcommand{\ml}{~\Msun ~\rm yr^{-1}}

\documentclass[preprint]{aastex}

\begin{document}

\title{YOUNG CIRCUMSTELLAR DISKS NEAR EVOLVED MASSIVE STARS AND SUPERNOVAE}
\author{Roger A. Chevalier}
\affil{Department of Astronomy, University of Virginia, P.O. Box 3818}
\affil{Charlottesville, VA 22903; rac5x@virginia.edu}


\begin{abstract}

There is increasing evidence that low mass stars with circumstellar disks
can be born close to massive stars, in some cases within tenths of a pc.
If the disks have lifetimes greater than those of the more massive stars,
they are exposed to the radiation fields and gas flows from the late evolutionary
phases and supernovae of the massive stars.
The fast flows from  supernovae are
likely to give some stripping of mass from the disks, but do not typically lead
to complete disruption of the disks.
In the slow wind from a red supergiant star, there is the possibility
of gravitational accretion of wind matter onto the circumstellar disk.
In the supernova explosion of a red supergiant, the radiative flux at the
time of shock breakout can heat and ionize a nearby disk, leading to
transient, narrow line emission.
There are consequences for the solar nebula if it was born $\sim 0.2$ pc from
a massive star that became a red supergiant.
Accretion from the wind could supply a substantial amount of $^{26}$Al
to the disk. 
The high radiative luminosity of the eventual supernova 
could lead to the melting of dust grains
and the formation of chondrules.
The passage of the supernova ejecta could drive a shock wave in the disk,
heating it.

\end{abstract}

\keywords{solar system: formation --- stars: circumstellar matter}

\section{INTRODUCTION}

There is growing evidence that low mass stars form in association with
massive stars.
Infrared studies of Orion show evidence for a cluster of about 2,000
low mass stars within 2 pc of the Trapezium; the core
radius of the cluster is $\sim 0.2$ pc (Hillenbrand \& Hartmann 1998).
Stars in Orion within $0.3$ pc of the central star $\theta^1$ C Ori show
observable disks   because of the effects of photoionization
by the massive star (O'Dell, Wen, \& Hu 1993).
The disk ages may be limited by
mass loss  driven  by a photodissociation front 
(St\"orzer \& Hollenbach 1999).
However, the age of the association of $\sim 10^6$ years and the fact that
 disks  are apparently present within
$\sim 0.01$ pc of $\theta^1$ C Ori  shows that the disks can survive,
perhaps because their host stars are on radial orbits 
(St\"orzer \& Hollenbach 1999).
Absorption studies of $\pi$ Sco in the 
upper Sco association give evidence for a population of circumstellar
disks that have properties like those in Orion (Bertoldi \& Jenkins 1992).
The estimated age of the upper Sco association is $5\times 10^6$ years
for both low and high mass stars,
with a small age dispersion 
(de Geus 1992; Preibisch \& Zinnecker 1999).
de Geus (1992) presents evidence that a supernova occurred in the association
$(1-1.5)\times 10^6$ years ago; it is possible that these circumstellar
 disks have been
exposed to a nearby supernova.

There have been estimates of the lifetime of the solar nebula
 $\sim 10^6$ years, but Podosek \& Cassen (1994) argue that a lifetime
$\sim 10^7$ years or more is  plausible.
These longer lifetimes, together with evidence for clusters of
low mass pre-main sequence stars near massive stars, suggests that
many gaseous disks may be exposed to the late evolutionary phases and
supernovae of massive stars.
The lifetime of a $60\Msun$ (initial mass),
 or O5 main sequence, star is $\sim 4\times 10^6$
years and that of a $20\Msun$ (O9) star
 is $\sim 1\times 10^7$ years.
In their late phases, these stars go through a red supergiant and/or a
Wolf-Rayet phase, with strong stellar winds.
They end their lives as supernovae.

In view of the fact that many circumstellar disks may be exposed to
these unusual environments, the aim of this paper is to investigate their
effects on the disks.
This is of interest both for  possible observational consequences
of the disks and for possible implications for the early solar system, if
it went through such a phase.
The effects of radiation on disks and of flows past a disk 
are discussed in \S~2 and \S~3, respectively.
The possible implications for the early solar system are in \S~4.

\section{EFFECTS OF RADIATION}

The effects of a main sequence stellar radiation field on a disk  
 have
been treated in detail for the case of Orion.
St\"orzer \& Hollenbach (1999) find that the FUV (far ultraviolet) stellar
flux determines the mass loss through a photodissociation front.
In this regime the mass loss rate does not have a strong dependence
on the stellar radiation flux.
However, once the radiation flux drops below the flux at which the FUV emission
dominates, the EUV (extreme ultraviolet) emission dominates and the mass loss
rate drops with the stellar flux.
For the case of $\theta^1$ C Ori, St\"orzer \& Hollenbach (1999) estimate
that the transition occurs at a distance of $\sim 0.3$ pc from the O star,
which is the outer limit at which disks are observed.
The rate of ionizing photon production, $S_*$, from $\theta^1$ C Ori is
$\sim 2.6\times 10^{49}$ s$^{-1}$, within the range of $2\times 10^{48}$
to $8\times 10^{49}$ s$^{-1}$ for O9 to O4 main sequence stars (Panagia 1973).
The Orion case may thus be typical of expectations around an O star.

Similar conditions are relevant to the case of Wolf-Rayet stars,
if that is the next phase of evolution of the massive star.
The case of $\gamma^2$ Velorum is of interest because it is the
brightest Wolf-Rayet star and X-ray observations show evidence for
a surrounding cluster of pre-main sequence stars (Pozzo et al. 2000).
There is no evidence at present for disks around these stars, the presence of
which would be important for establishing disk lifetimes.
For a distance of 410 pc, these stars extend to a radial distance
from $\gamma^2$ Velorum of 2 pc.
$\gamma^2$ Velorum is a WC8 star in a binary system with an O9 I star
(Conti \& Smith 1972).
The ultraviolet radiation field is dominated by the O9 I star, which has
$S_*\approx 1.3\times 10^{49}$ s$^{-1}$ (Panagia 1973).
Although this is comparable to the case of $\theta^1$ C Ori, the stronger
wind expected for the Wolf-Rayet star could affect the appearance
of nebulae around  disks (see \S~3).

When the massive star becomes a supernova, the highest luminosity
 occurs when the shock front breaks out of the stellar
surface and there is a  burst of hard radiation.
In the case of a red supergiant progenitor, $\sim 2\times 10^{48}$ ergs of
hard ultraviolet/soft X-ray radiation is emitted over $10^3$
s (Klein \& Chevalier 1978; Matzner \& McKee 1999).
For a Type Ib or Ic supernova from a
Wolf-Rayet star, the smaller surface area leads to
$3\times 10^{44}-2\times 10^{46}$ ergs of
soft X-ray radiation  emitted over $2-20$
s (Matzner \& McKee 1999).
The case of the unusual supernova SN 1987A is intermediate between these.
The initial burst can be followed by prolonged X-ray emission if the
shock wave runs into a dense circumstellar wind.

The supernova radiation can heat and ionize gas in the disk, leading
to a delayed burst of line radiation.
I consider a disk with mass $M_d=0.01 M_{d-2}\Msun$ and radius $d=
10^{15}d_{15}$ cm, so that the typical surface density is
$\sigma_d=M_d/(\pi d^2)=6.4 M_{d-2}d_{15}^{-2}$ g cm$^{-2}$.
The disk scale height at a radius $0.5 d$ is
$H\approx 0.7\times 10^{14} d_{15}^{3/2}T_2^{1/2}(M_*/\msun)^{-1/2}$ cm,
where $T$ is the temperature of the disk in units of $10^2$ K
and $M_*$ is the mass of the host star,
so that the typical density in the disk is 
$\rho_d\approx 1\times 10^{-14}
d_{15}^{-7/2}T_2^{-1/2}M_{d-2}(M_*/\msun)^{1/2}$ g cm$^{-3}$.
For the explosion of a red supergiant with a $2\times 10^{48}$ erg initial
burst of radiation, the amount of radiative energy
intercepted by a disk is $1\times 10^{42}(r/0.2{\rm~pc})^{-2}$ ergs,
where $r$ is the distance of the disk from the massive star.
At $r=0.2$ pc, the supernova radiation is absorbed in the outer layers
of the disk and reradiated; if $n_H=5\times 10^{9}$ cm$^{-3}$, an ionized layer 
$2\times 10^{12}$ cm thick can come into equilibrium with the
ionizing luminosity.
If present, dust grains would compete with gas for the ionizing photons,
but they are likely to be evaporated by the high luminosity (see below).
Based on calculations by Chevalier \& Fransson (1994), I estimate that
several \% of the reradiated luminosity is in the H$\alpha$ line.
The timescale for the emission at high density is determined by the light travel
time, $\sim 7\times 10^3$ s, leading to an H$\alpha$ luminosity of
$6\times 10^{36}(r/0.2{\rm~pc})^{-2}$ ergs s$^{-1}$ which could be observed
up to $1.3(r/0.2{\rm~pc})$ yr after the initial explosion.

Although this emission is faint compared to the supernova luminosity,
there is some chance of detection because it is in a narrow line.
The case we have been able to observe in most detail is SN 1987A,
which apparently had a $20\Msun$ progenitor star with a lifetime
of $1\times 10^7$ years (Arnett et al. 1989).
This is close to the age limit at which substantial disks are
expected to be present.
Efremov (1991) finds that the supernova is at the edge of a star cluster with
the expected age, but it is sparse (see also Panagia et al. 2000).
A number of short-lived, discrete sources of H$\alpha$ emission were
observed near the supernova in the first two years after the explosion
(Cumming \& Meikle 1993 and references therein). 
Although its parameters ($d\approx 2\times 10^{15}$ cm and
$r\approx 0.5$ pc) are reasonable for a circumstellar disk,
the best observed emission knot (Cumming \& Meikle 1993) appears unlikely
to be a  disk because of 
the $11\kms$ redshift of the narrow H$\alpha$ line compared to
the centroid of the bright ring emission.
Cumming \& Meikle (1993) estimate a density 
 of $(1-2)\times 10^5$ cm$^{-3}$ from the 60 day timescale of the emission;
 this is low for a circumstellar disk, but  it may be that
only the surrounding, low density parts of the disk were heated and ionized.
SN 1987A was a compact and relatively low luminosity supernova so that
dust should survive at this distance from the supernova, but the
dust opacity through the ionized region is not expected to be large.
Cumming \& Meikle (1993) find that the energy in the H$\alpha$ line was
consistent with the expected ionizing radiation from the supernova.

The supernova radiation can  process dust in the disk (by
vaporization and melting), and the surviving dust can emit scattered light
as well as  reradiate infrared light from absorbed supernova radiation.
Vaporization is caused by the highest luminosity radiation that is 
absorbed by dust.
From the results of Dwek (1983) for $n=1$, where the dust opacity
is $\kappa\propto\lambda^{-n}$, and an evaporation temperature of
$T_v=1500$ K, the dust is evaporated out to 
$r_v\approx 0.3 L_{45}^{1/2}$ pc from the supernova, where $L_{45}$ is the
luminosity in units of $10^{45}\ergs$.
The initial ionizing burst has $L_{45}\approx 2$ so that a disk at $r=0.2$
pc has its dust withing the ionized region evaporated.
The luminosity rapidly drops, so that a layer just interior to
the evaporated region is expected to have its
dust melted.
The properties of this layer depend on the disk position and
the grain properties.

Dust in circumstellar disks can reprocess the non-ionizing supernova
light, which typically has a total energy $\sim 2\times 10^{49}$ ergs
and $L_{45}<0.01$ for a
Type II supernova, to
 scattered and reradiated infrared light.
The problem with detecting this emission
is that it is not transient and is likely to
be difficult to distinguish from dust emission from the presupernova
wind (Dwek 1983).
The disks are optically thick to the supernova light, so the fraction
of the supernova radiative energy that is reprocessed depends on the
area covering factor of circumstellar disks.
Due to the initial dust evaporation,
the massive star wind is likely to be optically thin to the supernova light, but
it has a 100\% covering factor.

\section{EFFECTS OF FLOWS}

In addition to strong radiation fields, the environment of an
evolved massive star can have flows which affect a nearby circumstellar disk.
In the case of main sequence stars,  the properties of the cometary
nebulae observed around disks in Orion can be explained by the effects of radiation
from the massive star (e.g., McCollough et al. 1995;
St\"orzer \& Hollenbach 1999).
Although there is evidence for a stand-off bow shock in the wind from
$\theta^1$ C Ori in some cases (McCollough et al. 1995), the wind does
not play a role in driving mass loss from the disks.
The mass loss rate from  Wolf-Rayet stars is typically 100 times larger
than that from  massive main sequence stars.
Their wind velocities are comparable so that the wind ram pressure
is 100 times larger for the Wolf-Rayet case at a given distance from the
stars.
If the radiatively driven flows are comparable in the two cases, the 
wind bow shock is then 10 times closer to the disk in the Wolf-Rayet case,
at a radius of $\sim 10^{15}$ cm,
which places it at about the same radius as the photoionization front
for nebulae like those in Orion (McCollough et al. 1995).
The wind can thus affect the optical appearance of the nebula, by making it
somewhat more compact.

For a strong, fast flow, like that of a nearby supernova, matter
may be stripped from a disk.
If the timescale for the flow interaction is longer than the dynamical timescale
for the disk, the ram pressure of the flow can come into equilibrium with
the gravitational forces maintaining the disk.
If the interaction is rapid, the question is whether the momentum in the flow
can cause disk material to reach escape velocity.
The escape velocity from the outer parts of the disk is
$v_{esc}=(2GM_*/d)^{1/2}=5.2(M_*/\msun)^{1/2}d_{15}^{-1/2}\kms$, where
where $G$ is the gravitational constant.
The disk dynamical timescale is $t_d=d/v_{esc}=
61(M_*/\msun)^{-1/2}d_{15}^{3/2}$ yr.

In ram pressure stripping, the disk is disrupted if the ram pressure
in the flow, $\rho_f v_f^2$ where $\rho_f$ is the local density in
the flow and $v_f$ is the velocity, exceeds the gravitational force per 
unit area, $p_{grav}$, that keeps the 
disk bound to the central pre-main sequence star.
An estimate of the gravitational force per unit area is $
p_{grav}\approx GM_* \sigma_d/ d^2\approx 
2\times 10^{-3}\left(M_*/ \msun\right)
M_{d-2} d_{15}^{-4}
{\rm~dynes~cm^{-2}}.   $

The supernova case is complicated by the uncertainties in the density
distributions of the supernova and the surrounding medium.
In order to obtain an estimate of the effects, I assume a constant
density supernova with ejecta mass $M_{ej}$ and
energy $E$ in a constant density medium with H density $n_0$
(assuming a 10:1 H to He ratio by number).
The ram pressure effect is largest if the disk is directly exposed to
the freely expanding supernova ejecta, i.e. it becomes placed within
the reverse shock wave of the supernova remnant.
For a constant density supernova, the maximum radius of the reverse
shock is $  
r_{rs}=4.4\left(M_{ej}/ 10\Msun\right)^{1/3} n_0^{-1/3}{\rm~pc} $
(Truelove \& McKee 1999).
For the values of $r$ considered here, the
 circumstellar disk is exposed to the expanding
ejecta.
The maximum ram pressure that can be exerted by the ejecta occurs if 
the disk is hit by the outer, undecelerated edge of the ejecta, which
moves with a velocity $v_e=(10E/3M_{ej})^{1/2}=4.1\times 10^3
E_{51}^{1/2}(M_{ej}/10\Msun)^{-1/2}\kms$, where $E_{51}$ is in
units of $10^{51}$ ergs.
The supernova interaction time is $t_{SN}\approx r/v_e
=240(r/{\rm pc})E_{51}^{-1/2}(M_{ej}/10\Msun)^{1/2}$ yr, which is equal
to the disk dynamical timescale, $t_d$, for $r\approx 0.25$ pc.
For $r\ga 0.25$ pc, the ram pressure stripping arguments are
relevant, but for a close supernova, the momentum in the ejecta is
the important factor.
The peak ram pressure is $
p_{ram}=5E/ (2\pi r^3)=3\times 10^{-5}
E_{51}
(r/{\rm pc})^{-3} {\rm~dynes~cm^{-2}},  $
which is to be compared to $p_{grav}$ above.
A disk at 1 pc from the supernova can survive even this extreme case.
At  distances smaller than 0.25 pc, momentum transfer causes stripping to occur,
beginning at the outer parts of the disk.
The criterion for stripping is now that $M_{ej}v_{e}/(4\pi r^2)
> \sigma_d v_{esc}$.
It can be seen that the ram pressure and momentum stripping criteria are
roughly the same when the age equals the disk dynamical time,
or $r/v_{e}\approx d/v_{esc}$.

Stripping is the most likely consequence of a strong flow past a disk, 
but there may be circumstances
under which some of the flow is accreted to the disk.
The problem with the fast winds from main sequence or Wolf-Rayet stars
or with supernova ejecta is that the wind gas is shock 
heated to a high temperature
($10^7 [v_f/1000\kms]^2$ K for a H rich gas) and the radiative cooling time
is longer than the flow time.
Unless the gas can mix with the cool disk gas and share heat with it,
it is implausible that fast flow gas can be added to a disk.
The slow wind from a red supergiant star can marginally cool in a flow time
and gravitational effects can be significant for its accretion.
The Bondi-Hoyle accretion rate for a supersonic flow is $\dot M\approx 
4\pi(GM_*/v_f^2)^2\rho_w v_f$, where $\rho_w$ is the local density in
the wind and $v_f$ is the wind velocity.
Integrating over the time of accretion, the 
total accreted mass is $2\times 10^{-9}
(M_*/M_{\sun})^2(v_f/10{\kms})^{-4} (r/{\rm pc})^{-2} M_l$, 
where $M_l$ is the total mass
lost during the red supergiant phase.

\section{THE EARLY SOLAR SYSTEM}

In view of the evidence for the birth of low mass stars close to
massive stars, it is interesting to speculate on the consequences for
the proto-solar system if it were born is such an environment.
St\"orzer \& Hollenbach (1999) note that the disks close to
$\theta^1$ C Ori ($r\la 0.3$ pc) can have their outer parts 
 photoevaporated by the strong radiation field from
the star.
They further note that if this applied to the solar system, it 
could help explain why Uranus and Neptune have considerably less hydrogen
than Jupiter and Saturn.

There are further consequences of  the solar system being
born at $\sim 0.2$ pc from a massive star.
When the massive star became a red supergiant, at an age of $(3-4)\times 10^6$
years, the solar system would have
been enveloped by a slow dense wind.
Gravitational accretion of the wind would lead to $\sim 1.4\times 10^{-7}\Msun$
of dusty gas being accreted on the solar system, if the massive star lost
about $3\Msun$ of material during this phase.
The late accretion of grains would contribute matter that would not
have to pass through the possibly destructive environment of the
early solar nebula formation, although the accretion process could lead
to grain destruction.
The wind material would contain radioactive $^{26}$Al, which is known
to have been present at the birth of the solar system (Lee et al. 1976).
The abundance of $^{26}$Al in the wind from the H envelope
is $2\times 10^{-6}$ by mass for a $25\Msun$ star (Meyer, Woosley,
\& Weaver 1995),
so that $3\times 10^{-13}\Msun$ of $^{26}$Al might be accreted.
This is less than the $3\times 10^{-9}\Msun$ of $^{26}$Al that is required if
the $^{26}$Al was mixed throughout the proto-solar system including the sun
(Cameron et al. 1995), but it is sufficient to contaminate
$1\times 10^{-4}\Msun$ of disk gas to the observed level. 
The initial mass of the protoplanetary disk must be $\ga 10^{-2}\Msun$, but
only part of it may contain $^{26}$Al.
Other extinct radioactivities 
(e.g., Cameron et al. 1995 and references therein) are 
produced deeper in the massive star
and are not ejected in the red supergiant wind unless there is
a mechanism to mix them into the outer layers.
The inner layers would be ejected past the solar system during the
Wolf-Rayet and/or supernova phases, but I have argued above that the
high velocity of the gas in these phases makes accretion difficult.
If the gas could stick, the picture would resemble the early ``fly-paper model''
 of T. Gold (see p. 267 in Clayton 1977).

As discussed in \S~3, a layer of melted grains may be formed soon after
the time of supernova shock wave breakout, depending on the disk
position and grain properties.
This is of interest for solar system chondrules, which are grains
that have been melted (Hewins 1997).
In the supernova case, the grains would be rapidly heated by the
arrival of radiation at the time of shock breakout, but would 
cool more slowly because of the continued supernova radiation.
There is some evidence that the chondrules were rapidly heated
and subsequently cooled on a longer timescale (Hewins 1997).
The determination of whether supernova radiation is a suitable heat
source will require more detailed calculations.

The supernova gas would have reached the solar nebula tens of years
after the explosion.
If the disk had already been limited to a size of 10 a.u. by the
earlier action of photoevaporation, the passage of the ejecta would
not further disrupt the disk, but it would drive a shock front into the disk.
The ratio of $p_{grav}$ discussed above to the thermal pressure in
the disk  $p_{grav}/p_{th}\approx v_{esc}/c_d$ is typically $>10$, 
where $c_d$
is the sound speed in the disk.
The pressure due to ejecta approaches $p_{grav}$ and  can drive
a shock wave.
The shock wave propagation depends on the detailed density distribution in
the disk.
Shock waves in the solar nebula have been a leading explanation for the
formation of chondrules (Boss 1996) and provide an alternative to the
radiative heating.
If cooling of the shocked material is slow, some ablation of the disk gas
is possible.
The supernova shock wave in the red supergiant wind
generates a high pressure region
that envelops the circumstellar disk for tens of years.
For a wind mass loss rate of $3\times 10^{-6}\ml$, $v_f=10\kms$, and
a shock velocity of $v_{sh}=4,000\kms$, the pressure is
$\rho_w v_{sh}^2=6\times 10^{-6}$ dynes cm$^{-2}$, $\sim 10^7$ times the
present interstellar pressure.
The cosmic ray energy density is likely to be similarly enhanced because
of shock acceleration of particles; relativistic electrons 
in such shocked layers are observed
 in the radio supernova phenomenon.
Cosmic rays have been suggested as a source of isotopic anomalies in
meteorites (e.g., Clayton \& Jin 1995), but the energy available in the
current situation does not appear to be sufficient for interesting effects.

This speculative scenario of coeval formation of the solar system
and a nearby massive star is an alternative to the hypothesis that the
formation of the solar system was triggered by a supernova and that
radioactive isotopes were injected at that time (e.g., Cameron et al.
1995; Foster \& Boss 1996, 1997).
A major difference is that
in the present scenario, the early solar system is considerably closer to
the supernova and is exposed to a more extreme environment.
It is able to survive because of the gravitational binding of a
disk to its central star.
The unusual environment has a bearing on several perplexing properties
of solar system material.
The probability that the solar system went through such a phase is small
but perhaps non-negligible.
If the Orion Nebula Cluster has a limiting radius of 2 pc, $\sim 0.1$ of the
solar mass stars are within 0.2 pc of the center (Hillenbrand \& Hartmann
1998).
The cluster is likely to become sparser as is ages (Hillenbrand \& Hartmann
1998), so the probability of a nearby massive star may decline by
the time of a supernova.
This probability will be better determined by further observations of
circumstellar disks in massive star clusters and by studies aimed at
finding disks near  massive stars
in their final evolutionary phases.
The proposal made here of a search for transient narrow H$\alpha$ line emission
in the spectra of Type II supernovae is one possibility, although it is only
near the more massive supernovae ($\ga 20\Msun$ initial mass) that young
disks are likely to be present.

\acknowledgments
I am grateful to Donald Clayton and Zhi-Yun Li for comments on the
manuscript and to the referee for an especially useful report.
Support for this work was provided in part by NASA grant NAG5-8088.


\clearpage



\begin{thebibliography} {}

\bibitem[]{}
Arnett, W. D., Bahcall, J. N., Kirshner, R. P., \& Woosley, S. E. 1989,
ARA\&A, 27, 629

\bibitem[]{}
Bertoldi, F., \& Jenkins, E. B. 1992, ApJ, 388, 495


\bibitem[]{}
Boss, A. P. 1996, in Chondrules and the Protoplanetary Disk, ed. R. H. Hewins,
R. H. Jones, \& E. R. D. Scott (Cambridge: CUP), 257


\bibitem[]{}
Cameron, A. G. W., H\"oflich, P., Myers, P. C., \& Clayton, D. D.
1995, ApJ, 447, L53

\bibitem[]{}
Chevalier, R. A., \& Fransson, C. 1994, ApJ, 420, 268

\bibitem[]{}
Clayton, D. D. 1977, Icarus, 32, 255

\bibitem[]{}
Clayton, D. D., \& Jin, L. 1995, ApJ, 451, L87

\bibitem[]{}
Cumming, R. J., \& Meikle, W. P. S. 1993, MNRAS, 262, 689

\bibitem[]{}
Conti, P. S., \& Smith, L. F. 1972, \apj, 172, 623

\bibitem[]{}
Dwek, E. 1983, \apj, 274, 175

\bibitem[]{}
Efremov, Y. N. 1991, Sov. Astr. Lett., 17, 173

\bibitem[]{}
de Geus, E. J. 1992, \aap, 262, 258

\bibitem[]{}
Foster, P. N., \& Boss, A. P. 1996, ApJ, 468, 784

\bibitem[]{}
Foster, P. N., \& Boss, A. P. 1997, ApJ, 489, 336


\bibitem[]{}
Hewins, R. H. 1997, Ann. Rev. Earth Plan. Sci., 25, 61

\bibitem[]{}
Hillenbrand, L., \& Hartmann, L. W. 1998, \apj, 499, 758



\bibitem[]{}
Klein, R. I., \& Chevalier, R. A. 1978, \apj, 223, L109

\bibitem[]{}
Lee, T., Papanastassiou, D. A., \& Wasserburg, G. J. 
1976, Geophys. Res. Lett., 3, 41


\bibitem[]{}
Matzner, C. D., \& McKee, C. F. 1999, \apj, 510, 379


\bibitem[]{}
McCullough, P. R., Fugate, R. Q., Christou, J. C., Ellerbroed, B. L.,
Higgins, C. H., Spinhirne, J. M., Cleis, R. A., \& Moroney, J. F.
1995, \apj, 438, 394


\bibitem[]{}
Meyer, B. S., Weaver, T. A., \& Woosley, S. E. 1995, Meteoritics, 30, 325

\bibitem[]{}
O'Dell, C. R.,  Wen, Z., \& Hu, X. 1993, ApJ, 410, 696

\bibitem[]{}
Panagia, N. 1973, AJ, 78, 929

\bibitem[]{}
Panagia, N., Romaniello, M., Scuderi, S., \& Kirshner, R. P. 2000, ApJ, in press
(astro-ph/0001476)

\bibitem[]{}
Podosek, F. A., \& Cassen, P. 1994, Meteoritics, 29, 6

\bibitem[]{}
Pozzo, M., Jeffries, R. D., Naylor, T., Totten, E. J., Harmer, S.,
\& Kenyon, M. 2000, MNRAS, 313, L23  

\bibitem[]{}
Preibisch, T., \& Zinnecker, H. 1999, \aj, 117, 2381

\bibitem[]{}
St\"orzer, H., \& Hollenbach, D. 1999, ApJ, 515, 669

\bibitem[]{}
Truelove, J. K., \& McKee, C. F. 1999, ApJS, 120, 299

%


\end{thebibliography}
\end{document}